\documentstyle[11pt,epsfig]{article}
\title{\bf Gauss-Bonnet brane gravity with a confining potential}
\author{M. Heydari-Fard \thanks{email:
m-heydarifard@sbu.ac.ir} and H. R. Sepangi\thanks{email:
hr-sepangi@sbu.ac.ir}
\\ {\small Department of Physics, Shahid Beheshti University, Evin,
Tehran 19839, Iran}}
\begin{document}
\maketitle 

\begin{abstract}
A brane scenario is envisaged in which the $m$-dimensional bulk is
endowed with a Gauss-Bonnet term and localization of matter on the
brane is achieved by means of a confining potential. The resulting
Friedmann equations on the brane are modified by various extra
terms that may be interpreted as the X-matter, providing a
possible phenomenological explanation for the accelerated
expansion of the universe. The age of the universe in this
scenario is studied and shown to be consistent with the present
observational data.
\end{abstract}
\section{Introduction}
Recent years have been witnessing a phenomenal interest in the
possibility that our observable four-dimensional ($4D$) universe
may be viewed as a brane hypersurface embedded in a higher
dimensional bulk space. Physical matter fields are confined to
this hypersurface, while gravity can propagate in the higher
dimensional space-time as well as on the brane. The most popular
model in the context of brane world theory is that proposed by
Randall and Sundrum (RS). In the so-called RSI model \cite{1},
they proposed a mechanism to solve the hierarchy problem with two
branes, while in the RSII model \cite{2}, they considered a single
brane with a positive tension, where $4D$ Newtonian gravity is
recovered at low energies even if the extra dimension is not
compact. This mechanism provides us an alternative to
compactification of extra dimensions. The cosmological evolution
of such a brane universe has been extensively investigated and
interesting effects such as dark radiation or quadratic density
term in Friedmann equations have been found \cite{3}.

There is a general belief that Einstein gravity is the low-energy
limit of a quantum theory of gravity which is still unknown. A
promising candidate is the string theory which suggests that in
order to have a ghost-free action, quadratic curvature corrections
to the Einstein-Hilbert action must be proportional to the
Gauss-Bonnet (GB) term \cite{4}. This term also plays a
fundamental role in Chern-Simons gravitational theories \cite{5}.
The appearance of higher derivative gravitational terms can also
be seen in the renormalization of quantum field theories in curved
space-times \cite{6}. From a geometric point of view, the
combination of the Einstein-Hilbert and Gauss-Bonnet term
constitutes, for $5D$ space-times, the most general Lagrangian
producing second-order field equations \cite{7}. In four
dimensions, the Gauss-Bonnet combination reduces to a total
divergence and is dynamically irrelevant. These facts provide a
strong motivation for the study of braneworld theories which
include a Gauss-Bonnet term. Interestingly, it has been shown that
the graviton zero mode is as localized at low energies in the
Gauss-Bonnet brane system as in the RSII model \cite{8}-\cite{10}
and that the correction of the Newton's law becomes less
pronounced by including the Gauss-Bonnet term \cite{11}.

As the discussion above suggests, it comes as no surprise that
brane theories with a Gauss-Bonnet term have been generating so
much interest over the recent past \cite{12}. One of the results
of these investigations is that the Gauss-Bonnet correction to the
Friedmann equations has been dominant in the early universe,
having the form $H^2\propto\rho^{2/3}$ at high energies. This term
arises from the imposition of the generalized Israel junction
conditions \cite{13}. However, it has been argued that such
junction conditions may not be unique. Indeed, other forms of
junction conditions exist, so that different conditions may lead
to different physical results \cite{14}. Furthermore, these
conditions cannot be used when more than one non-compact extra
dimension is involved. To avoid such concerns, an interesting
higher-dimensional model was introduced in \cite{15} where
particles are trapped on a 4-dimensional hypersurface by the
action of the confining potential ${\cal V}$. In \cite{16}, the
dynamics of test particles confined to a brane by the action of
such potential at the classical and quantum levels were studied
and the effects of small perturbations along the extra dimensions
investigated. Within the classical limits, test particles remain
stable under small perturbations and the effects of the extra
dimensions are not felt by them, making them undetectable in this
way. The quantum fluctuations of the brane cause the mass of a
test particle to become quantized and, interestingly, the
Yang-Mills fields appear as quantum effects. Also, in \cite{17}, a
braneworld model was studied in which matter is confined to the
brane through the action of such a potential, rendering the use of
any junction condition unnecessary and predicting a geometrical
explanation for the accelerated expansion of the universe.

In brane theories the covariant Einstein equations are derived by
projecting the bulk equations onto the brane. This was first done
in \cite{18} where the Gauss-Codazzi equations together with
Israel junction conditions  were used to obtain the Einstein field
equations on the 3-brane. The same procedure was subsequently used
in \cite{20} to obtain the field equations on the brane in the
presence of the Gauss-Bonnet term  where to confine matter on the
brane, the authors used a delta-function in the energy-momentum
tensor. A feature of these works is that the effective field
equations are too complicated to be physically transparent because
the extrinsic curvature cannot be expressed solely by the
energy-momentum tensor on the brane. Still, they are important in
assessing the effects of the higher order terms appearing in the
action. It would therefore be interesting to investigate the
effects of such higher order terms when a confining potential,
rather than junction conditions, is used to localize matter on the
brane.

In this paper, we consider an $m$-dimensional bulk space in the
presence of the Gauss-Bonnet term without imposing the $Z_2$
symmetry \cite{21}. As mentioned above, to localize matter on the
brane, a confining potential is used rather than a delta-function
in the energy-momentum tensor. The resulting Friedmann equation is
modified by the appearance of extra terms that behave like the
X-matter; the phenomenological model proposed to fit the data
explaining accelerated expansion of the universe. Within the
framework of our model, we also calculate the age of the universe
and show that it is consistent with the bounds set by the modern
observational data. We should emphasize here that there is a
difference between the model presented in this work and models
introduced in \cite{9,10} in that in the latter no mechanism is
introduced to account for the confinement of matter to the brane.
\section{Geometrical considerations}
Consider the background manifold $ \overline{V}_{m-1} $
isometrically embedded in a pseudo-Riemannian manifold $ V_{m}$ by
the map ${ \cal Y} : \overline{V}_{m-1}\rightarrow  V_{m} $ such
that
\begin{eqnarray}\label{a}
{\cal G} _{AB} {\cal Y}^{A}_{,\mu } {\cal Y}^{B}_{,\nu}=
\bar{g}_{\mu \nu}  , \hspace{.5 cm} {\cal G}_{AB}{\cal
Y}^{A}_{,\mu}{\cal N}^{B} = 0  ,\hspace{.5 cm}  {\cal G}_{AB}{\cal
N}^{A}{\cal N}^{B} =\epsilon= \pm 1.\label{a}
\end{eqnarray}
where $ {\cal G}_{AB} $  $ ( \bar{g}_{\mu\nu} ) $ is the metric of
the bulk (brane) space  $  V_{m}  (\overline{V}_{m-1}) $ in
arbitrary coordinates, $ \{ {\cal Y}^{A} \} $   $  (\{ x^{\mu} \})
$  is the basis of the bulk (brane) and  ${\cal N}^{A}$ denotes
the vector normal to the brane. The perturbation of
$\bar{V}_{m-1}$ in a sufficiently small neighborhood of the brane
along an arbitrary transverse direction $\xi$ is given by
\begin{eqnarray}\label{b}
{\cal Z}^{A}(x^{\mu},\xi) = {\cal Y}^{A} + ({\cal L}_{\xi}{\cal
Y})^{A}, \label{b}
\end{eqnarray}
where $\cal L$ represents the Lie derivative. By choosing $\xi$
orthogonal to the brane, we ensure gauge independency \cite{maia1}
and have perturbations of the embedding along a single orthogonal
extra direction $\bar{{\cal N}}$ giving local coordinates of the
perturbed brane as
\begin{eqnarray}\label{c}
{\cal Z}^{A}_{,\mu}(x^{\nu},\xi) = {\cal Y}^{A}_{,\mu} +
\xi\bar{{\cal N}}^{A}_{,\mu}(x^{\nu}),\label{c}
\end{eqnarray}
where $\xi$ is a small parameter along ${\cal N}^{A}$ that
parameterizes the extra noncompact dimensions. One can see from
equation (\ref{b}) that since the vectors $\bar{{\cal N}}^{A}$
depend only on the local coordinates $x^{\mu}$, they do not
propagate along the extra dimensions
\begin{eqnarray}\label{d}
{\cal N}^{A}(x^{\mu}) = \bar{{\cal N}}^{A}+ \xi[\bar{{\cal N}} ,
\bar{{\cal N}}]^{A} = \bar{{\cal N}}^{A}.\label{d}
\end{eqnarray}
The embedding equations of the perturbed hypersurface $
\overline{V}_{m-1} $ is given by
\begin{eqnarray}\label{e}
{\cal G}_{AB}{\cal Z}_{\,\,\ ,\mu }^{A}{\cal Z}_{\,\,\ ,\nu
}^{B}={\cal G}_{\mu \nu },\hspace{0.5cm}{\cal G}_{AB}{\cal
Z}_{\,\,\,,\mu}^{A}{\cal N}^{B}=0,\hspace{0.5cm}{\cal G}_{AB}{\cal
N}^{A}%
{\cal N}^{B}=\epsilon= \pm 1.\label{e}
\end{eqnarray}
Denoting by $\bar{K}_{\mu \nu }=-{\cal G}_{AB}{\cal Y}_{\,\,\,,\mu
}^{A}{\cal N}_{;\nu }^{B}$ the extrinsic curvature of
$\overline{V}_{m-1} $, and using (\ref{c}), the perturbed metric
${g}_{\mu\nu}$ can be written as
\begin{equation}\label{f}
g_{\mu \nu }=\bar{g}_{\mu \nu }-2\xi \bar{K}_{\mu \nu }+\xi^{2}%
\bar{g}^{\alpha \beta }\bar{K}_{\mu \alpha }\bar{K}_{\nu \beta },
\label{f}
\end{equation}%
with the extrinsic curvature of the perturbed brane given by
\begin{equation}\label{g}
{K}_{\mu \nu }=-{\cal G}_{AB}{\cal Z}_{\,\,\,,\mu }^{A}{\cal
N}_{\,\,\,;\nu }^{B}=\bar{K}_{\mu\nu}- \xi \bar{K}_{\mu \alpha
}\bar{K}^{\alpha}_{\,\,\,,\nu}. \label{g}
\end{equation}
Comparison of equations (\ref{f}) and (\ref{g}) then leads to the
York relation, describing the metric evolution with respect to the
perturbation parameter $\xi$
\begin{equation}\label{h}
{K}_{\mu \nu }=-\frac{1}{2}\frac{\partial {\cal G}_{\mu \nu }}{%
\partial \xi}.  \label{h}
\end{equation}
The components of the Riemann tensor of the bulk written in the
the embedding $\{{\cal Z}^{A}, {\cal N}^A \}$  lead to the
Gauss-Codazzi equations given by \cite{24}
\begin{eqnarray}
{\cal R}_{ABCD}{\cal Z} ^{A}_{,\alpha}{\cal Z} ^{B}_{,\beta}{\cal
Z} ^{C}_{,\gamma} {\cal Z}^{D}_{,\delta}=R_{\alpha \beta \gamma
\delta}+2K_{\alpha[ \gamma }K_{\beta] \delta },\label{i}
\end{eqnarray}
\begin{equation}
{\cal R}_{ABCD}{\cal Z} ^{A}_{,\alpha} {\cal N}^{B} {\cal Z}
^{C}_{,\gamma} {\cal
Z}^{D}_{,\delta}=K_{\beta\gamma;\alpha}-K_{\alpha\gamma;\beta},\label{j}
\end{equation}
together with \cite{20}
\begin{equation}
{\cal R}_{ABCD} ~{\cal Z}_{~,\mu}^{A}\, {\cal Z} _{~,\nu}^{C}\,
{\cal N}^{B}\,{\cal N} ^D= -{\cal L}_{\cal N} K_{\mu\nu}
+K_{\mu\alpha}\, K^{\alpha}_{~\nu}\,\label{k},
\end{equation}
where ${\cal R}_{ABCD}$ and $R_{\alpha\beta\gamma\delta}$ are the
Riemann tensors for the bulk and the perturbed brane respectively
and ${\cal L_{N}}$ denotes the lie derivative in the ${\cal
N}$-direction. Using this projection, the ${m}$-dimensional
Riemann curvature and its contractions (the Ricci tensor and
scalar curvature) are described by the $(m-1)$-dimensional
variables on the brane with the normal ${\cal N}_A$ as \cite{20}
\begin{eqnarray}\label{l}
{\cal R}_{ABCD} &=&R_{ABCD}-K_{AC}K_{BD}+K_{AD}K_{BC} -{\cal N}_A
K_{BD;C}+{\cal N}_A K_{BC;D}+{\cal N}_B K_{AD;C} -{\cal N}_B
K_{CA;D} \nonumber
\\
&& -{\cal N}_C K_{BD;A}+{\cal N}_C K_{AD;B}+{\cal N}_D
K_{BC;A}-{\cal N}_D K_{AC;B} \nonumber
\\
&& +{\cal N}_A{\cal N}_C K_{BE}K^E_{~D} -{\cal N}_A{\cal N}_D
K_{BE}K^E_{~C} -{\cal N}_B{\cal N}_C K_{AE}K^E_{~D} +{\cal
N}_B{\cal N}_D K_{AE}K^E_{~C} \nonumber
\\
&&-{\cal N}_A{\cal N}_C {{\cal L}_{\cal N}} K_{BD} +{\cal
N}_A{\cal N}_D {{\cal L}_{\cal N}} K_{BC} +{\cal N}_B{\cal N}_C
{{\cal L}_{\cal N}} K_{AD} -{\cal N}_B{\cal N}_D {{\cal L}_{\cal
N}} K_{AC}, \label{l}
\end{eqnarray}
\begin{eqnarray}\label{ll}
{\cal R}_{AB} &=&R_{AB}-KK_{AB}+2K_{AC}K^C_{~B} +{\cal N}_A
\left(K^C_{~B;C}- K_{;B}\right) +{\cal N}_B \left( K^C_{~A;C}-
K_{;A}\right)
\nonumber\\
&& +{\cal N}_A{\cal N}_B K_{CD}K^{CD} - {{\cal L}_{\cal N}} K_{AB}
-{\cal N}_A{\cal N}_B g^{CD}{{\cal L}_{\cal N}} K_{CD},
 \label{ll}
\end{eqnarray}
\begin{eqnarray}\label{lll}
{\cal R} &=&R-K^2+3K_{CD}K^{CD} -2 g^{CD}{{\cal L}_N}
K_{CD}.\label{lll}
\end{eqnarray}
The Einstein-Gauss-Bonnet equation is quasi-linear \cite{11},
which means that apart from non-singular terms given by the
$(m-1)$-dimensional variables, it contains only linear terms of
${{\cal L}_{\cal N}} K_{AB}$  with no quadratic terms appearing.
\section{Effective field equations}
We consider the total action for space-time $(\cal M,{\cal
G}_{AB})$ with boundary $(\Sigma,g_{\mu\nu})$ as
\begin{eqnarray}\label{mm}
S=\frac{1}{2\alpha^{*}}\int_{\cal M} d^{m}X \sqrt{-\cal G} ({\cal
R}-2\Lambda^{(b)}+\beta{\cal L}_{GB})+\int_{\Sigma} d^{m-1}x
\sqrt{-g} ({\cal L}_{\rm surface}+{\cal L}_{m}).
\end{eqnarray}
Variation of the total action gives our basic field equations as
\begin{eqnarray}\label{m}
{G}^{(b)}_{AB} +\Lambda^{(b)}{\cal G}_{AB}+\beta {\cal
H}^{(b)}_{AB}=\alpha^{*} S_{AB},\label{m}
\end{eqnarray}
where
\begin{eqnarray}\label{n}
{G}^{(b)}_{AB}&=&{\cal R}_{AB}-{1\over 2}{\cal G}_{AB}{\cal R},
\\
{\cal H}^{(b)}_{AB}&=&2\left[{\cal R}{\cal R}_{AB}-2{\cal
R}_{AC}{\cal R}^C_{~B} -2{\cal R}^{CD}{\cal R}_{ACBD}+ {\cal
R}_A^{~CDE}{\cal R}_{BCDE}\right]-{1\over 2}{\cal G}_{AB}{\cal
L}_{GB}, \label{n}
\end{eqnarray}
and
\begin{eqnarray}\label{o}
{\cal L}_{GB}={\cal R}^2-4{\cal R}_{AB}{\cal R}^{AB} +{\cal
R}_{ABCD}{\cal R}^{ABCD}\label{o}
\end{eqnarray}
is the Gausss-Bonnet term. In the above equation
$\alpha^{*}=\frac{1}{M_{*}^{m-2}}$ ($M_{*}$ is the fundamental
scale of energy in the bulk space), ${\beta}$ is the Gauss-Bonnet
coupling constant with dimension $(\mbox{length})^2$,
$\Lambda^{(b)}$ is the cosmological constant of the bulk and
$S_{AB}$ consists of two parts
\begin{equation}\label{p}
S_{AB}=T_{AB}+ \frac{1}{2} {\cal{V}} {\cal{G}}_{AB},\label{p}
\end{equation}
where $T_{AB}\equiv -2 \frac {\delta {\cal L}_{\rm m} }{ \delta
g^{AB}}  +g_{AB}{\cal L}_{\rm m}$ is the energy-momentum tensor of
the matter confined to the brane through the action of the
confining potential $\cal{V}$. We require $\cal{V}$  to satisfy
three general conditions: firstly, it has a deep minimum on the
non-perturbed brane, secondly, depends only on extra coordinates
and thirdly, the gauge group representing the subgroup of the
isometry group of the bulk space is preserved by it \cite{16}.
Substituting relations (\ref{l}), (\ref{ll}) and (\ref{lll}) into
Eq. (\ref{m}), we find the effective equations on the brane as
\begin{eqnarray}\label{q}
P_{\mu\nu}&-&{1\over
2}Pg_{\mu\nu}+K_{\mu\rho}K^\rho_{~\nu}-g_{\mu\nu}K_{\alpha\beta}
K^{\alpha\beta}-{{\cal L}_{\cal N}}
K_{\mu\nu}+g_{\mu\nu}g^{\rho\sigma} {{\cal L}_{\cal N}}
K_{\rho\sigma}
\nonumber \\
&+&2\beta\left(H_{\mu\nu} -P{{\cal L}_{\cal N}} K_{\mu\nu}
+2P^{~\rho}_{\mu} {{\cal L}_{\cal N}} K_{\rho\nu}
+2P^{~\rho}_{\nu} {{\cal L}_{\cal N}} K_{\rho\mu}
+W_{\mu\nu}^{~~~\rho\sigma} {{\cal L}_{\cal N}} K_{\rho\sigma}
\right)\nonumber\\ &=&{\alpha}^{*} {S}_{AB}{\cal Z}_{~,\mu}^A
{\cal Z}_{~,\nu}^B-\Lambda^{(b)}
g_{\mu\nu} \,, \label{qq}\\
& &P+\beta\left(P^2-4P_{\alpha\beta}P^{\alpha\beta}
+P_{\alpha\beta\gamma\delta}P^{\alpha\beta\gamma\delta}\right)
=-2{\alpha}^{*}{S}_{AB} {\cal N}^A{\cal N}^B
+2\Lambda^{(b)},\label{qqqq}
\end{eqnarray}
where
\begin{eqnarray}\label{r}
P_{\alpha\beta\gamma\delta} &=&
R_{\alpha\beta\gamma\delta}-K_{\alpha\gamma}K_{\beta\delta}
+K_{\alpha\delta}K_{\beta\gamma},
\nonumber \\
P_{\alpha\beta}
&=&g^{\rho\sigma}P_{\alpha\rho\beta\sigma}=R_{\alpha\beta}
-KK_{\alpha\beta}+K_{\alpha\gamma}K^{\gamma}_{~\beta},
\nonumber \\
P
&=&g^{\alpha\beta}P_{\alpha\beta}=R-K^2+K_{\alpha\beta}K^{\alpha\beta},
\\
H_{\mu\nu} &=&
PP_{\mu\nu}-2(P_{\mu\rho}P^{\rho}_{~\nu}+P^{\rho\sigma}
P_{\mu\rho\nu\sigma}
)+P_{\mu\rho\sigma\kappa}P_{\nu}^{~\rho\sigma\kappa}
+2K_{\alpha\beta}K^{\alpha\beta}P_{\mu\nu}+
PK_{\mu\rho}K^\rho_{~\nu}\nonumber \\
&-&2(K_{\mu\rho}K^\rho_{~\sigma}P^\sigma_{~\nu}
+K_{\nu\rho}K^\rho_{~\sigma}P^\sigma_{~\mu}) -2
K^{\rho\kappa}K_\kappa^{~\sigma}P_{\mu\rho\nu\sigma}
\nonumber \\
&-&{1\over 4}g_{\mu\nu}\left[
P^2-4P_{\alpha\beta}P^{\alpha\beta}+P_{\alpha\beta\gamma\delta}
P^{\alpha\beta\gamma\delta}\right]
\nonumber \\
&+&g_{\mu\nu}\left[-K_{\alpha\beta}K^{\alpha\beta}P
+2P_{\alpha\beta}K^{\alpha\gamma}K_{\gamma}^{~\beta} \right],
  \\
W_{\mu\nu}^{~~~\rho\sigma} &=&
Pg_{\mu\nu}g^{\rho\sigma}-2P_{\mu\nu}g^{\rho\sigma}-2g_{\mu\nu}
P^{\rho\sigma}
+2P_{\mu\alpha\nu\beta}g^{\alpha\rho}g^{\beta\sigma} \,.\label{r}
\end{eqnarray}
In order to find the effective equations on the brane, we have to
replace the  terms ${{\cal L}_{\cal N}} K_{\mu\nu}$ in Eq.
(\ref{q}) with  the $(m-1)$-dimensional variables on the brane.
Using the decomposition of the Riemann tensor into the Weyl
curvature, the Ricci tensor and the scalar curvature
\begin{eqnarray}\label{s}
{\cal R}_{ABCD}=C_{ABCD}-\frac{2}{(m-2)}\left({\cal G}_{A[D}{\cal
R}_{C]B}-{\cal G}_{B[D}{\cal
R}_{C]A}\right)-\frac{2}{(m-1)(m-2)}{\cal R}({\cal G}_{A[C}{\cal
G}_{D]B}),\label{s}
\end{eqnarray}
we obtain
\begin{eqnarray}\label{t}
{{\cal L}_{\cal N}} K_{\mu\nu}-{1\over (m-1)} g_{\mu\nu}
g^{\alpha\beta}{{\cal L}_{\cal N}} K_{\alpha\beta}&=&-{(m-2)\over
(m-3)} {\cal E}_{\mu\nu}+K_{\mu\rho}K^{\rho}_{~\nu} -{1\over
(m-3)}\left[ P_{\mu\nu} -{1\over
(m-1)}Pg_{\mu\nu}\right]\nonumber\\ &-&{1\over (m-1)}
g_{\mu\nu}K_{\rho\sigma}K^{\rho\sigma}\,, \label{t}
\end{eqnarray}
where
\begin{equation}\label{u}
{\cal E}_{\mu\nu}={\cal C}_{ABCD}~{\cal N}^B {\cal N}^D {\cal
 Z}_{~,\mu}^{A}~ {\cal
 Z} _{~,\nu}^{C}, \label{u}
\end{equation}
is the electric part of the Weyl tensor ${\cal C}_{ABCD}$.
However, since  Eq. (\ref{t}) is a trace free equation, we cannot
fix ${{\cal L}_{\cal N}} K_{\mu\nu}$ by it. We have to find
$g^{\alpha\beta}{{\cal L}_{\cal N}} K_{\alpha\beta}$ from the
other independent equation. If we take the trace of our basic Eq.
(\ref{m}), we find
\begin{eqnarray}\label{v}
(m-2){\cal R} +\beta(m-4){\cal
L}_{GB}=-2\alpha^{*}{S}+2m\Lambda^{(b)}. \label{v}
\end{eqnarray}
Substituting Eqs. (\ref{i})-(\ref{k}) with Eq. (\ref{t}) into Eq.
(\ref{v}), we obtain
\begin{eqnarray}\label{w}
g^{\alpha\beta}{{\cal L}_{\cal N}} K_{\alpha\beta}={P\over
2}+K_{\alpha\beta}K^{\alpha\beta}+{{(\alpha^{*}S-m\Lambda^{(b)})}\over
\left[(m-2)+\beta P(m-4)\right]}
 +{\beta(m-4)I\over 2\left[(m-2)+\beta
P(m-4))\right]} , \label{w}
\end{eqnarray}
where
\begin{eqnarray}\label{x}
I= P^2-8P_{\alpha\beta}P^{\alpha\beta}+P_{\alpha\beta\gamma\delta}
P^{\alpha\beta\gamma\delta}-12P_{\rho\sigma}{\cal
E}^{\rho\sigma}.\label{x}
\end{eqnarray}
From Eq. (\ref{t}) with Eq. (\ref{w}), we then obtain
\begin{eqnarray}\label{A}
{{\cal L}_{\cal N}} K_{\mu\nu}&=&{(m-2)\over (m-3)} {\cal
E}_{\mu\nu} -{1\over (m-3)}\left(P_{\mu\nu}-{1\over
2}Pg_{\mu\nu}\right) +{(\alpha^{*} S-m\Lambda^{(b)})\over
(m-1)\left[(m-2)+\beta P(m-4)\right]} g_{\mu\nu}
\nonumber \\
&&+K_{\mu\rho}K^{\rho}_{\nu}+{(m-4)\beta I\over
2(m-1)\left[(m-2)+\beta P(m-4)\right]} g_{\mu\nu}. \label{A}
\end{eqnarray}
Substituting Eq. (\ref{A}) into Eq. (\ref{q}), we obtain the
effective gravitational equations on the brane as
\begin{eqnarray}\label{B}
&& {(m-2)\over (m-3)}\left(P_{\mu\nu}+{\cal E}_{\mu\nu}\right)
+\beta\left(H^{(1)}_{\mu\nu}+H^{(2)}_{\mu\nu}
\right)-{2(m-4)^2\beta^2I\over (m-1)\left[(m-2)+\beta
P(m-4)\right]}\left( P_{\mu\nu}-{1\over
4}Pg_{\mu\nu}\right)\nonumber\\
&&-{P g_{\mu\nu}\over 2(m-3)}=\alpha^{*}
S_{\mu\nu}-\Lambda^{(b)}g_{\mu\nu} +\beta{4(m-4)\over
(m-1)}{{(\alpha^{*}S-m\Lambda^{(b)})}\over \left[(m-2)+\beta
P(m-4)\right]}\left(P_{\mu\nu}-{1\over 4}P
g_{\mu\nu}\right)\nonumber\\
&&-{{(\alpha^{*}S-m\Lambda^{(b)})}\over (m-1)}g_{\mu\nu} \,,
\label{B}
\end{eqnarray}
where
\begin{eqnarray}\label{C}
H^{(1)}_{\mu\nu}
&=&2P_{\mu\alpha\beta\gamma}P_{\nu}^{~\alpha\beta\gamma}-{4(m-2)\over
(m-3)}P^{\rho\sigma} P_{\mu\rho\nu\sigma}+{6\over (m-3)}P
P_{\mu\nu}-{4(m-2)\over (m-3)}P_{\mu\rho}P_{\nu}^{~\rho}
\nonumber \\
&+&{g_{\mu\nu}\over
2(m-1)(m-3)}\left[(m^2-9m+6)P^2+4(m^2-4m+7)P_{\alpha\beta}P^{\alpha\beta}\right.\nonumber\\
&-&\left.
\left(m^2-5m+6\right)P_{\alpha\beta\gamma\delta}P^{\alpha\beta\gamma\delta}\right],
\label{cc}
\\H^{(2)}_{\mu\nu} &=&-{4(m-2)\over
(m-3)}\left(P_{\mu\rho}{\cal E}^\rho_{~\nu}+P_{\nu\rho}{\cal
E}^\rho_{~\mu}+ P_{\mu\rho\nu\sigma}{\cal E}^{\rho\sigma}
\right)\nonumber\\ &+&{2(m^2-7)\over
(m-1)(m-3)}g_{\mu\nu}P_{\rho\sigma}{\cal E}^{\rho\sigma}
+{2(m-2)\over (m-3)}P{\cal E}_{\mu\nu}.\label{C}
\end{eqnarray}
As was mentioned in the introduction, localization of matter on
the brane is realized  in this model by the action of a confining
potential. Let us take
\begin{eqnarray}\label{F}
\alpha\tau_{\mu\nu}={\alpha^{*}(m-3)\over(m-2)}T_{\mu\nu},\hspace{.5
cm} \Lambda=-{(m-3)\over(m-1)}\Lambda^{(b)},\label{F}
\end{eqnarray}
where $\alpha$ is the scale of energy on the brane. Now, we
rewrite Eq. (\ref{B}) as an Einstein-type equation with
``correction" terms. From Eqs. (\ref{qqqq}) and (\ref{B}) we
obtain
\begin{eqnarray}\label{D}
G_{\mu\nu}&+&\beta\left(\hat{H}^{(1)}_{\mu\nu}+\hat{H}^{(2)}_{\mu\nu}
\right)=\alpha \tau_{\mu\nu}+\Lambda g_{\mu\nu
}-\frac{\alpha\tau}{(m-1)}g_{\mu\nu}+Q_{\mu\nu}-{\cal E}_{\mu\nu}
\nonumber \\
&+&{\beta
(P_{\mu\nu}-\frac{1}{4}Pg_{\mu\nu})\over\left[(m-2)+\beta
P(m-4)\right]} \left[{4(m-4)\over(m-1)}\alpha
\tau+{4m(m-4)\over(m-2)}\Lambda\right], \label{D}
\end{eqnarray}
where
\begin{eqnarray}\label{G}
\hat{H}^{(1)}_{\mu\nu}&=&{2(m-3)\over(m-2)}
P_{\mu\alpha\beta\gamma}
P_{\nu}^{~\alpha\beta\gamma}-4P^{\rho\sigma}
P_{\mu\rho\nu\sigma}+{6\over(m-2)}PP_{\mu\nu}-4P_{\mu\rho}P_{\nu}^{~\rho}
\nonumber \\
&-&{g_{\mu\nu}\over
2(m-1)(m-2)}\left[(-5m+3)P^2-4(2m^2-8m+10)P_{\alpha\beta}P^{\alpha\beta}\right.\nonumber\\
&+&\left.
(-2m^2+9m-9)P_{\alpha\beta\gamma\delta}P^{\alpha\beta\gamma\delta}\right]
-{2\beta\over \left[(m-2)+\beta
P(m-4)\right]}{(m-4)^2(m-3)\over(m-1)(m-2)}\nonumber\\
&\times &\left(P_{\mu\nu}-{1\over 4}Pg_{\mu\nu}\right)
\left(P^2-8P_{\alpha\beta}P^{\alpha\beta}
+P_{\alpha\beta\gamma\delta}P^{\alpha\beta\gamma\delta}\right),
\\
\hat{H}^{(2)}_{\mu\nu}&=&-4\left(P_{\mu\rho}{\cal E}^\rho_{~\nu}
+P_{\nu\rho}{\cal E}^\rho_{~\mu}+ P_{\mu\rho\nu\sigma}{\cal
E}^{\rho\sigma} \right)
+{2(m^2-7)\over(m-1)(m-2)}g_{\mu\nu}P_{\rho\sigma}{\cal
E}^{\rho\sigma}+2P{\cal E}_{\mu\nu} \nonumber\\
&+&{24\beta \over \left[(m-2)+\beta
P(m-4)\right]}{(m-4)^2(m-3)\over(m-1)(m-2)}
\left(P_{\mu\nu}-{1\over 4}Pg_{\mu\nu}\right)P_{\rho\sigma}{\cal
E}^{\rho\sigma}. \label{GG}
\end{eqnarray}
Here, $\Lambda$ is the effective cosmological constant in four
dimensions with $Q_{\mu\nu}$ being a completely geometrical
quantity given by
\begin{eqnarray}\label{E}
Q_{\mu\nu}&=&KK_{\mu\nu}-K_{\mu\rho}K^\rho_{~\nu}+{1\over 2}
\left( K_{\alpha\beta}K^{\alpha\beta}-K^2
\right)g_{\mu\nu}.\label{E}
\end{eqnarray}

A brief discussion on the energy-momentum conservation on the
brane would be in order here. The contracted Bianchi identities in
the bulk space $G^{AB(b)}_{\,\,\,\,\,\,\,\,\,;A}=0$ and ${\cal
{H}}^{AB(b)}_{\,\,\,\,\,\,\,\,\,;A}=0$, using Eq. (\ref{m}), imply
\begin{eqnarray}
\left(T^{AB}+\frac{1}{2} {\cal{V}} {\cal {G}}^{AB}\right)_{
;A}=0.\label{eq1}
\end{eqnarray}
Since the potential $\cal V$ has a minimum on the brane, the above
conservation equation reduces to
\begin{eqnarray}\label{a22}
\tau^{\mu\nu}_{\,\,\,\,\,;\mu}=0,
\end{eqnarray}
As a check of our calculations so far, we note that Eq. (\ref{D})
reduces to Eq. (32) given in \cite{17} in a 5-dimensional bulk
with no Gauss-Bonnet term, that is $\beta=0$.  As can be seen from
the definition of $Q_{\mu\nu}$, it is independently a conserved
quantity which, according to \cite{21}, may be interpreted as an
energy-momentum tensor of a dark energy fluid representing the
$X$-matter, also known as the `X-Cold-Dark Matter' (XCDM). This
matter has the most general form of the equation of state which is
characterized by the following conditions \cite{22}: first it
violates the strong energy condition at the present epoch for
$\omega_x<-1/3$ where $p_x=\omega_x\rho_x$, second, it is locally
stable {\it i.e.} $c^2_s=\delta p_x/\delta\rho_x\ge 0$, and third,
causality holds good, that is $c_s\le 1$. Ultimately, we have
three different types of ``matter'' on the right hand side of Eq.
(\ref{D}), namely, ordinary confined conserved matter represented
by $\tau_{\mu\nu}$, the matter represented by $Q_{\mu\nu}$ which
will be discussed later and finally, the Weyl matter represented
by ${\cal E}_{\mu\nu}$.
\section{Modified Friedmann equations}
In this section we will analyze the influence of the trace of
$\tau_{\mu\nu}$, the extrinsic curvature terms and higher
curvature terms on a FRW universe, regarded as a brane embedded in
an $m$-dimensional bulk with a Gauss-Bonnet term. The FRW line
element is given by
\begin{eqnarray}\label{1}
ds^2=-dt^2+a(t)^2\left[\frac{dr^2}{1-kr^2}+r^2\left(d\theta^2+\sin^2\theta
d\varphi^2\right)\right].
\end{eqnarray}
The confined source is the perfect fluid given in co-moving
coordinates by
\begin{eqnarray}\label{2}
\tau_{\mu\nu}=(\rho+p) u_{\mu}u_{\nu}+p g_{\mu\nu},\hspace{.5
cm}u_{\mu}=-\delta^{0}_{\mu},\hspace{.5 cm}p=(\gamma-1)\rho.
\end{eqnarray}
Let us consider an $AdS_m$, $dS_m$ or flat bulk, so that ${\cal
E}_{\mu\nu}=0$. For late times this assumption seems reasonable
because the effects of such a term is negligible. The Codazzi
equations (\ref{j}) reduce to
\begin{eqnarray}\label{3}
K_{\alpha\gamma ;\sigma}-K_{\alpha\sigma ;\gamma}=0.
\end{eqnarray}
Using the Yorks relation, it follows that in the FRW space-time
(diagonal metric), $K_{\mu\nu}$ is diagonal. After separating the
spatial components, the Codazzi equations reduce to
\begin{eqnarray}\label{4}
K_{\mu\nu ,\rho}-K_{\nu\sigma }\Gamma^{\sigma}_{\mu\rho}=
K_{\mu\rho ,\nu}-K_{\rho\sigma }\Gamma^{\sigma}_{\mu\nu},
\end{eqnarray}
\begin{eqnarray}\label{5}
K_{\mu\nu ,0}-K_{\mu\nu
}\frac{\dot{a}}{a}=-a\dot{a}\left(\delta^{1}_{\mu}\delta^{1}_{\nu}+r^2
\delta^{2}_{\mu}\delta^{2}_{\nu}+r^2\sin\theta^2\delta^{3}_{\mu}\delta^{3}_{\nu}\right)K_{00}.
\end{eqnarray}
The first equation gives $K_{11 ,\nu}=0$  for $\nu\neq1$, since
$K_{11 }$ does not depend on the spatial coordinates. After
defining $K_{11 }=b(t)$, where $b(t)$ is an arbitrary function of
$t$, the second equation gives
\begin{eqnarray}\label{5}
K_{00}=-\frac{1}{\dot{a}}\frac{d}{dt}\left(\frac{b}{a}\right).
\end{eqnarray}
For $\mu,\nu=2,3$ we obtain $K_{22}=b(t)r^2$ and
$K_{33}=b(t)r^2\sin^2\theta$ and generally $(\mu,\nu \neq 0)$
\begin{eqnarray}\label{6}
K_{\mu\nu}=\frac{b}{a^2}g_{\mu\nu}.
\end{eqnarray}
We find from Eq. (\ref{E}) that
\begin{eqnarray}\label{7}
Q_{\mu\nu}=-\frac{1}{a^4}\left(2\frac{b\dot{b}}{H}-b^2\right)g_{\mu\nu}
,\hspace{.5 cm} Q_{00}=\frac{3 b^{2}}{a^4}.
\end{eqnarray}
Denoting $h=\frac{\dot{b}}{b}$ and $H=\frac{\dot{a}}{a}$, the
components of $Q_{\mu\nu}$ become
\begin{eqnarray}\label{8}
Q_{\mu\nu}=-\frac{b^2}{a^4}\left(2\frac{h}{H}-1\right)g_{\mu\nu}
,\hspace{.5 cm} Q_{00}=\frac{3b^2}{a^4}.
\end{eqnarray}
It would now be interesting to see how the above geometrical
interpretation is compared with the X-matter explanation. To this
end we define $Q_{\mu\nu}$ as a perfect fluid and write
\begin{eqnarray}\label{9}
Q_{\mu\nu}=\frac{1}{\alpha}\left[(\rho_{x}+p_{x})
u_{\mu}u_{\nu}+p_{x} g_{\mu\nu}\right] ,\hspace{.5 cm}
p_{x}=(\gamma_{x}-1)\rho_{x}.
\end{eqnarray}
Comparison with the components of $Q_{\mu\nu}$ given by Eq.
(\ref{8}) gives
\begin{eqnarray}\label{10}
P_{x}=-\frac{1}{\alpha}\frac{b^2}{a^4}\left(2\frac{h}{H}-1\right)
,\hspace{.5 cm}\rho_{x}=\frac{3}{\alpha}\frac{b^2}{a^4}.
\end{eqnarray}
Use of the above equations leads to an equation for $b(t)$
\begin{eqnarray}\label{11}
\frac{\dot{b}}{b}=\frac{1}{2}\left(4-3\gamma_{x}(t)\right)\frac{\dot{a}}{a}.
\end{eqnarray}
It is interesting to note that this equation resembles one of the
phenomenological candidates for dark energy, the X-matter
\cite{22}, but in our case this field has a fundamental
geometrical justification for the equation of state, having been
derived from the term $Q_{\mu\nu}$ in the Einstein equation
(\ref{D}), itself a result of the extrinsic curvature. If
$\gamma_{x}$ is taken as a constant, the solution for $b(t)$ is
\begin{eqnarray}\label{12}
b(t)=b_{0}a(t)^{\frac{1}{2}(4-3\gamma_{x})},
\end{eqnarray}
where $b_{0}$ is an integration constant. Using Eq. (\ref{10}) and
this solution, the energy density of XCDM becomes
\begin{eqnarray}\label{13}
\rho_{x}=\frac{3b_{0}^{2}}{\alpha}a^{-3\gamma_{x}}.
\end{eqnarray}
It should be noted that we have considered an $AdS_m$, $dS_m$ or
flat bulk and consequently ${\cal E}_{\mu\nu}=0$, therefore Eq.
(\ref{eq1}) leads to $\hat{H}^{(2)}_{\mu\nu}=0$.

Now, using these relations and Eq. (\ref{D}), the Friedmann
equations, to first order in $\beta$ become
\begin{eqnarray}\label{14}
3H^{2}+f(a)+\beta\left[H^{2} A(a)+\dot{H}
B(a)+18H^{4}+4\dot{H}^{2}+16\dot{H}H^{2}+\hat{f}(a)\right]+{\cal
O}(\beta^{2})=0,
\end{eqnarray}
\begin{eqnarray}\label{15}
-2\dot{H}-3H^{2}+g(a)+\beta\left[H^{2} C(a)+\dot{H}
D(a)-18H^{4}-\frac{8}{3}\dot{H}^{2}-\frac{56}{3}\dot{H}H^{2}+\hat{g}(a)\right]+{\cal
O}(\beta^{2})=0,
\end{eqnarray}
where
\begin{eqnarray}\label{a}
A{(a)}&=&4\Lambda+\frac{20k}{a^2}-12b_{0}^2
a^{-3\gamma_{x}}(3-2\gamma_{x})-\alpha\rho_{0}a^{-3\gamma}(3\gamma),
\nonumber\\
B(a)&=&\frac{16\Lambda}{3}+\frac{8k}{a^2}-4b_{0}^2
a^{-3\gamma_{x}}(4-3\gamma_{x})-2\alpha\rho_{0}a^{-3\gamma},
\nonumber\\
f(a)&=&\Lambda+\frac{3k}{a^2}-3b_{0}^2
a^{-3\gamma_{x}}-\frac{3}{4}\alpha\rho_{0}a^{-3\gamma},
\nonumber\\
\hat{f}(a)&=&-(4-8{\gamma_{x}})\Lambda b_{0}^2 a^{-3\gamma_{x}}
+3b_{0}^4a^{-6\gamma_{x}}\left(+3\gamma_{x}^2-8\gamma_{x}+6\right)
+3\alpha\rho_{0}b_{0}^2a^{-3(\gamma+\gamma_{x})}(\gamma-\gamma_{x})\nonumber\\
&+&6\left(\frac{k}{a^2}\right)^2-\frac{4\Lambda
k}{3a^2}+k\alpha\rho_{0}(2-3\gamma)a^{-3\gamma-2}-kb_{0}^2(20-12\gamma_{x})a^{-3\gamma_{x}-2},
\label{a}
\end{eqnarray}
and
\begin{eqnarray}\label{aa}
C(a)&=&-4\Lambda-\frac{52k}{3a^2}+4b_{0}^2
a^{-3\gamma_{x}}(9-7\gamma_{x})-\alpha\rho_ {0}\gamma
a^{-3\gamma},
\nonumber\\
D(a)&=&-\frac{8\Lambda}{9}-\frac{40k}{3a^2}+4b_{0}^2
a^{-3\gamma_{x}}\left(\frac{14}{3}-2\gamma_{x}\right)-\frac{2}{3}\alpha\rho_{0}a^{-3\gamma},
\nonumber\\
g(a)&=&-\Lambda-\frac{k}{a^2}+3b_{0}^2
a^{-3\gamma_{x}}(1-\gamma_{x})-\frac{\alpha\gamma}{4}\rho_{0}a^{-3\gamma},
\nonumber\\
\hat{g}(a)&=&-
b_{0}^4a^{-6\gamma_{x}}\left(6\gamma_{x}^2-28\gamma_{x}+18\right)
+\alpha\rho_{0} b_{0}^2a^{-3(\gamma+\gamma_{x})}
(\gamma-\gamma_{x})+\Lambda b_{0}^2
a^{-3\gamma_{x}}\left(4-\frac{4\gamma_{x}}{3}\right)
\nonumber\\
&-&2\left(\frac{k}{a^2}\right)^2-\frac{28\Lambda
k}{9a^2}+k\alpha\rho_{0}\left(\frac{2}{3}-\gamma\right)a^{-3\gamma-2}
+kb_{0}^2\left(\frac{52}{3}-20\gamma_{x}\right)a^{-3\gamma_{x}-2}.
\end{eqnarray}\label{aa}
Equations (\ref{14}) and (\ref{15}) now constitute our  Fridmann
equations, having been modified by a  number of extra terms. Such
terms may be used to offer an explanation for the X-matter
discussed earlier. In the next section, we discuss the
implications of such terms on the cosmology of our model.
\section{Cosmological implications}
Recent observational data indicate that our universe is expanding
with a positive acceleration. This acceleration is explained in
terms of the so-called dark energy which could result from some
exotic form of matter with negative pressure. The nature of such
dark energy constitutes an open and tantalizing question
connecting cosmology and particle physics. The simplest form of
dark energy is the vacuum energy (cosmological constant). However,
this scenario faces the two well known cosmological and
coincidence problems \cite{cosmological constant}. Another
possible form of dark energy is provided by scalar fields. Dark
energy can be attributed to the dynamics of a scalar field, called
the quintessence \cite{quintessence}. Also a phenomenological
explanation based on current observational data is given by the
X-matter model which is associated with an exotic fluid
characterized by an equation of state $p_{x}=w_{x}\rho_{x}$, where
the parameter $w_{x}$ can be a constant or more generally a
function of time \cite{XCDM}. In this section, following the
method developed in \cite{17}, we show that within the context of
the present model, it is possible to have a universe exhibiting
accelerated expansion without the need to resort to a cosmological
constant term or a scalar field or any other kind of dark energy.

Let us proceed by taking equations (\ref{14}) and (\ref{15}) from
which, the Hubble parameter can be written as
\begin{eqnarray}\label{1a}
H^{2}&=&\frac{1}{210}\left[2\Lambda-\frac{210k}{a^2}+210 b_{0}^2
a^{-3\gamma_{x}}-3\left(7-\frac{27 \gamma}{2}\right)\alpha\rho_{0}
a^{-3\gamma}\right]\nonumber\\
&-&\frac{3}{70\beta}\pm\frac{3}{70\beta}\left\{1+2\beta
\left[-8\Lambda+\left(\frac{4\gamma+7}{3}\right)
\alpha\rho_{0}a^{-3\gamma}\right]\right\}^{\frac{1}{2}}.
\end{eqnarray}
We note that for $\beta\longrightarrow 0$ the negative sign yields
a singular term in the above equation and, therefore, we select
the positive sign. A qualitative classification of the solutions
on the basis of different values of the parameter $\gamma_{x}$ can
be achieved without solving the equation. If we define the
effective potential
\begin{eqnarray}\label{2a}
V(a)&=&k-b_{0}^2
a^{-3\gamma_{x}+2}+\frac{\alpha\rho_{0}}{70}\left(7-\frac{27
\gamma}{2}\right)
a^{-3\gamma+2}\nonumber\\
&+&\frac{3
a^2}{70\beta}-\frac{3a^2}{70\beta}\left\{1+\frac{2}{3}\beta
\left({4\gamma+7}\right)
\alpha\rho_{0}a^{-3\gamma}\right\}^{\frac{1}{2}},
\end{eqnarray}
we then have
\begin{eqnarray}\label{3a}
\dot{a}^2+V(a)=0.
\end{eqnarray}
Now, it is possible to deduce the qualitative behavior of the
scale factor $a(t)$, keeping in mind that $\dot a^2$ should be
positive. This behavior is much dependent on the range of the
values that $\gamma_{x}$ can take. We distinguish the following
possibility for having an accelerating universe \cite{23}
\begin{equation}\label{4a}
0 <\gamma_{x}<\frac{2}{3}.
\end{equation}
We also note that to avoid imaginary values for the effective
potential, $\beta>0$. The behavior of the potential for
$k=0,\pm1$, is illustrated in figure 1. This figure shows that,
within the context of the present model, the universe exhibits
accelerated expansion in the case of a vanishing cosmological
constant with $\beta>0$. This case is what we expect from the
slope of the expansion in heterotic string theory. It should be
mentioned that Gauss-Bonnet theories with a negative coupling
constant have been suggested \cite{25} in the past where cosmic
acceleration can be achieved, but this is not well motivated by
string theory and has problems with stability \cite{26}.

The above equation cannot be solved in closed form. However, in
two extreme cases corresponding to small and large $a(t)$, exact
solutions can be easily found
\begin{equation}\label{5a}
a(t)=\left(\frac{3}{2}\right)^{\frac{2}{3}}\left(\frac{13\alpha\rho_{0}}{140}\right)^{\frac{1}{3}}t^\frac{2}{3},
\hspace{.5 cm} \mbox{for small} \hspace{2mm}a,
\end{equation}
and
\begin{equation}\label{6a}
a(t)=\frac{b_{0}^{2}}{4}t^{2}, \hspace{.5 cm}  \mbox{for large}
\hspace{2mm}a.
\end{equation}
The first  solution is of the Einstein de-Sitter type, while the
second represents an evidently inflationary of the power-law type.
This means that in our model the universe starts as decelerating
and finally ends up as accelerating. In the simplest FRW
cosmological models with a one-component fluid filling up the
universe such behavior is not possible. In the next section we
discuss the observational parameters of our model.
\begin{figure}
\centerline{\begin{tabular}{ccc}\epsfig{figure=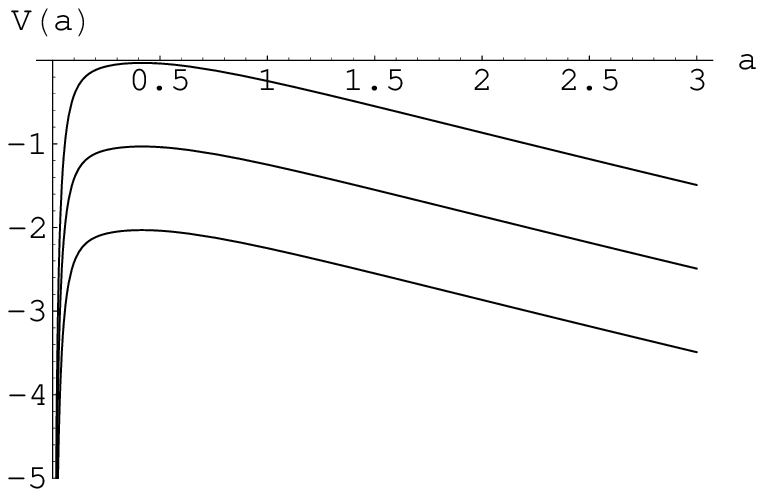,width=6cm}\hspace{20mm}
\epsfig{figure=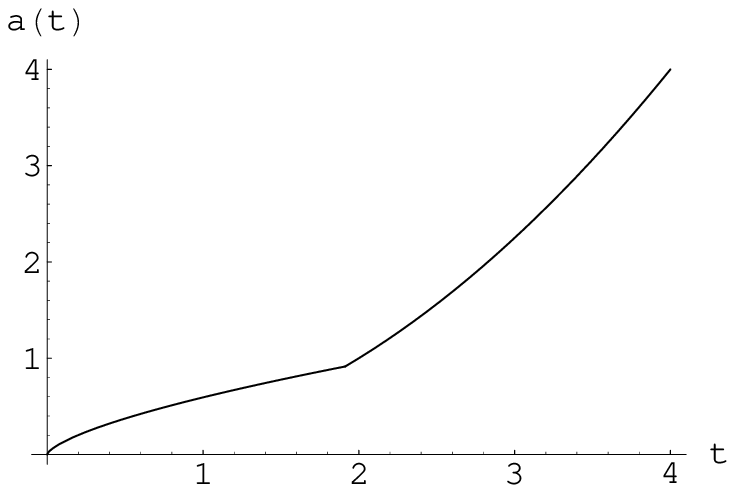,width=6cm}
\end{tabular}}
\caption{\footnotesize Left, the behavior of the potential $V(a)$
for $\gamma=1$, $\gamma_{x}=\frac{1}{3}$ and $k=0$ middle, $k=1$
top and $k=-1$ bottom and right, the behavior of the scale factor
as a function of $t$ for the same parameters and $k=0$ .}
\label{fig1}
\end{figure}
\section{Age of the universe}
The age of the universe in FRW models is given by
\begin{equation}\label{0b}
t_{0}^{F}=\frac{1}{H_{0}}\int_{0}^{1}\frac{dx}{\left[\frac{\Omega_{m}}{x}+
(1-\Omega_{m})\right]^\frac{1}{2}},
\end{equation}
where $H_{0}^{-1}=9.8 \times 10^9 \mbox{h}^{-1}$ years and the
dimensionless parameter h, according to modern data, is about 0.7.
Hence, in the flat matter dominated universe with
$\Omega_{total}=1$ the age of the universe would be only 9.3 Gyr,
whereas the oldest globular clusters yield an age of about 12.5
with an uncertainly of 1.5 Gyr \cite{27}. With the assumption of
the cosmological constant, the WMAP data \cite{28} yields
$t_{0}=13.7\pm0.2$ Gyr, and the best estimate of the dynamical age
of the universe, coming from CMB data, is $t_{0}=14\pm0.5$ Gyr
\cite{29}.

It would now be interesting to compare the above results and those
predicted by the present work to the results obtained in \cite{17}
where the same mechanism for localization of matter is used  in
the absence of the Gauss-Bonnent term and the age of the universe
is given by
\begin{equation}\label{1b}
t_{0}^{B}=\frac{1}{H_{0}}\int_{0}^{1}\frac{dx}{\left(\frac{\Omega_{m}}{x}+
\Omega_{x}x\right)^\frac{1}{2}},
\end{equation}
for a flat, matter dominated universe with $\Omega_{m}=0.3$ and
$\Omega_{x}=0.7$. This leads to a prediction for the age of the
universe of about 12 Gyr. Here, with the GB term present, we
introduce the dimensionless cosmological parameters
\begin{equation}\label{2b}
\Omega_{m}=\frac{\alpha\rho_{0m}}{3H_{0}^{2}},\hspace{.5
cm}\Omega_{x}=\frac{\alpha\rho_{0x}}{3H_{0}^{2}},\hspace{.5
cm}\Omega_{GB}=\beta H_{0}^{2},
\end{equation}
where the subscript zero refers to the present values of the
cosmological quantities. The modified Friedmann equation
(\ref{1a}) with $\Lambda=0$ and $k=0$ is then given by
\begin{equation}\label{3b}
\frac{H^{2}}{H_{0}^{2}}=\Omega_{x}\left(\frac{a}{a_{0}}\right)^{-3\gamma_{x}}-\frac{3}{70}\left(7-\frac{27
\gamma}{2}\right)\Omega_{m}\left(\frac{a}{a_{0}}\right)^{-3\gamma}
-\frac{3}{70\Omega_{GB}}\left\{1-\left[1+2
\left({4\gamma+7}\right) \Omega_{GB}
\Omega_{m}\left(\frac{a}{a_{0}}\right)^{-3\gamma}\right]^{\frac{1}{2}}\right\}.
\end{equation}
This equation leads to the constraint equation that should be
satisfied amongst the cosmological parameters defined above when
$a=a_0$, that is
\begin{equation}\label{4b}
1=\Omega_{x}-\frac{3}{70}\left(7-\frac{27
\gamma}{2}\right)\Omega_{m}
-\frac{3}{70\Omega_{GB}}\left\{1-\left[1+2
\left({4\gamma+7}\right) \Omega_{GB}
\Omega_{m}\right]^{\frac{1}{2}}\right\},
\end{equation}
which reduces the number of independent $\Omega$ parameters. We
therefore find the age of the universe by direct integration of
the Friedmann Eq. (\ref{3b})
\begin{eqnarray}\label{5b}
t_{0}^{GB}=\frac{1}{H_{0}}\int_{0}^{1}\frac{dx}{\left\{\Omega_{x}x^{-3\gamma_{x}+2}-
\frac{3}{70}(7-\frac{27\gamma}{2})\Omega_{m}x^{-3\gamma+2}
-\frac{3x^2}{70\Omega_{GB}}\left[1-\left(1+2(4\gamma+7)\Omega_{GB}\Omega_{m}
x^{-3\gamma}\right)^\frac{1}{2}\right]\right\}^\frac{1}{2}}.
\end{eqnarray}
The results are summarized in table 1. It shows that the age of
the universe in our model is longer than the FRW model and brane
models without the Gauss-Bonnet term in the bulk action \cite{17}.
Noting that the dynamical age of the universe depends on the rate
of the expansion, we realize that in the presence of the GB term,
the universe accelerates faster than brane models without the GB
term. We have plotted the age of the universe in three models as a
function of the energy density parameter $\Omega_{m}$ in figure 2.
Figure 3 shows the age of the universe as a function of
$\Omega_{m}$, corresponding to the values of $\gamma_{x}$ and
$\gamma$. In figure 4 , we have plotted the age as a function of
two parameters, ($\gamma_{x}$,$\Omega_{m}$) and
($\gamma$,$\Omega_{m}$). Note that the age of the universe
increases with decreasing $\gamma_{x}$, $\gamma$ and $\Omega_{m}$.
\begin{center}
\begin{tabular}{cccccc} \hline\hline$\gamma_{x}$ &
$t_{0}H_{0}$ & $t_{0}$ (Gyr)
\\ \hline\hline
&       &       &       \\
$0.1$  & 1.061 & 14.54 \\
$0.2$  & 1.037 & 14.21 \\
$0.3$  & 1.010 & 13.83 \\
$0.4$  & 0.978 & 13.41 \\
$0.5$  & 0.942 & 12.91 \\
$0.6$  & 0.901 & 12.34 \\
\\
\hline\hline
\end{tabular}\vspace{2mm}\\
{\footnotesize Table 1: Age of the universe for various values of
$\gamma_x$.}
\end{center}
\vspace{2mm}\noindent\\

At this point, it would be appropriate to compare these results
with other brane world models where the Gauss-Bonnet term is
absent. A very noticeable and popular scenario in this connection
is the Randall-Sundrum brane model where the effects of the brane
parameters and dark energy on the age of the universe have been
studied. It has been shown that the effect of the quadratic term
$\rho^2$, resulting from the imposition of the Israel junction
conditions, in the energy density term is to lower significantly
the age of the universe. The quadratic term contribution in the
energy density even for a small negative pressure contributes
effectively as the positive pressure, and makes brane models less
accelerating. This problem can be avoided if we accept the dark
energy $p=-\frac{4}{3}\rho$ (phantom matter) on the brane, since
it has a very strong influence on increasing the age \cite{30}.

Another scenario which has attracted a considerable amount of
attention in recent years is the so-called DGP model, proposed by
Dvali, Gabadadze and Porrati (DGP) \cite{31} and generalized to
cosmology by Defayet \cite{32}. This proposal explains the
observed late time acceleration of the expansion of the universe
through a large scale modification of gravity coming from the
``leakage'' of gravity at large scales into an extra dimension,
without requiring a non-vanishing cosmological constant \cite{33}.
In other words, the bulk gravity sees its own curvature term on
the brane as the cosmological constant and accelerates the
universe. From the observational viewpoint, it has been shown that
such models are in agreement with the most recent cosmological
observations. For example, in \cite{33}, the authors show that
constraints from the SNe Ia+CMB data require a flat universe with
$\Omega_{m}=0.3$ and $\Omega_{r_{c}}=0.12$, where $\Omega_{r_{c}}$
is the density parameter associated with the crossover distance
between the $4D$ an $5D$ gravities. Also, the age of the universe
in the context of these models have been studied and shown that
for a fixed value of $\Omega_{m}$, the predicated age  is longer
for larger values of $\Omega_{r_ {c}}$ and therefore these models
are more efficient in addressing the problem of estimating the age
of the universe \cite{34}.

\begin{figure}
\centerline{\begin{tabular}{ccc}
\epsfig{figure=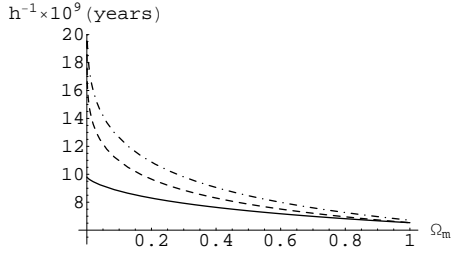,width=6cm}
\end{tabular} }
\caption{\footnotesize The age of universe with the GB term
(dot-dashed line), without the GB term (dashed line) and as
predicted by the FRW model (solid line), as a function of
$\Omega_{m}$ for$\gamma=1$, $\gamma_{x}=\frac{1}{3}$ and
$\Omega_{GB}=10^{-3}$. } \label{fig2}
\end{figure}
\begin{figure}
\centerline{\begin{tabular}{ccc}
\epsfig{figure=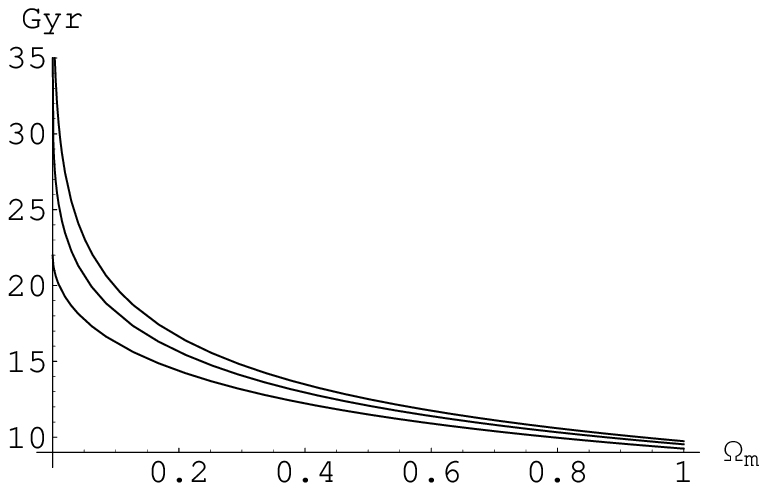,width=6cm}\hspace{20mm}\epsfig{figure=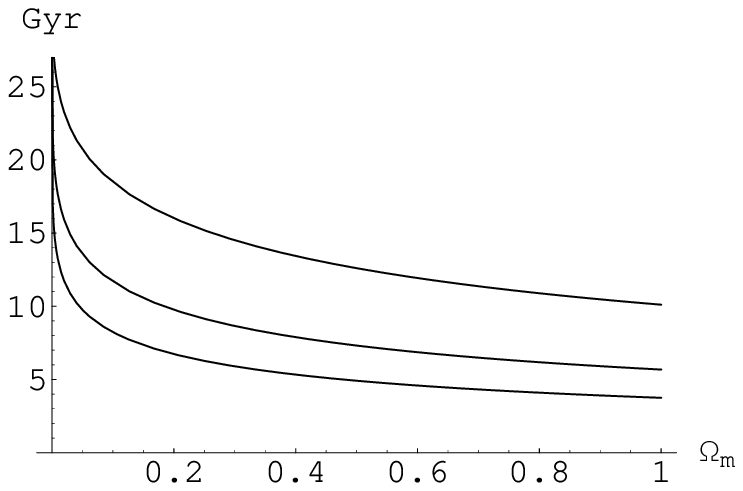,width=6cm}
\end{tabular} }
\caption{\footnotesize Left, the age of universe as predicted by
the present model as a function of $\Omega_{m}$ for different
values of $\gamma_{x}=0.5,0.3,0.1$, bottom, middle and top curves
respectively with $\gamma=1$ and $\Omega_{GB}=10^{-3}$ and right,
the age of universe as a function of $\Omega_{m}$ for different
values of $\gamma=2,1.5,1$, bottom, middle and top curves
respectively with $\gamma_{x}=\frac{1}{3}$ and
$\Omega_{GB}=10^{-3}$.} \label{fig3}
\end{figure}
\begin{figure}
\centerline{\begin{tabular}{ccc}
\epsfig{figure=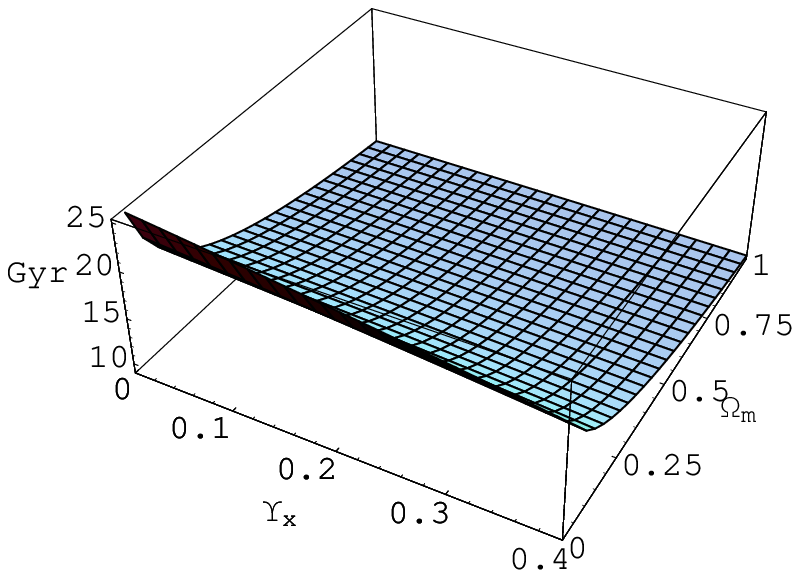,width=6cm}\hspace{20mm}\epsfig{figure=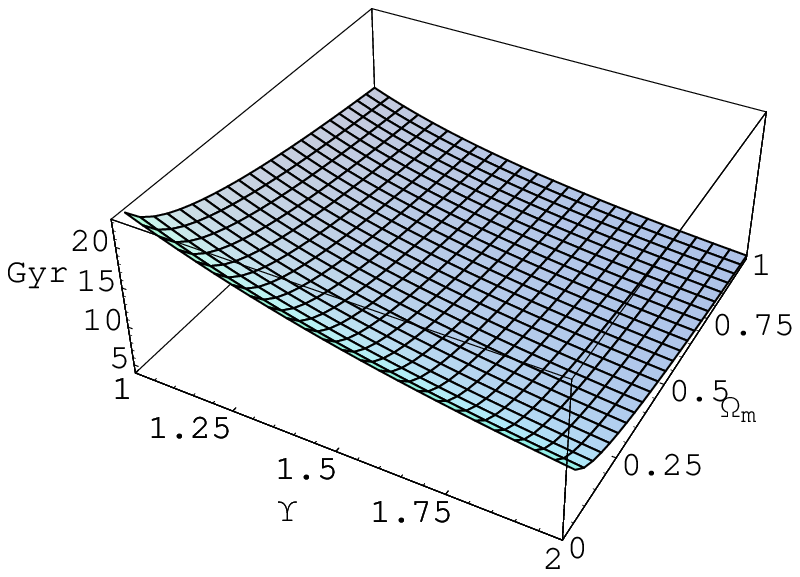,width=6cm}
\end{tabular} }
\caption{\footnotesize Age of the universe as predicted by the
present model as a function of $\gamma_{x}$ and $\Omega_{m}$, left
and as a function of $\gamma$ and $\Omega_m$, right. We have taken
$\Omega_{GB}=10^{-3}$. } \label{fig4}
\end{figure}
\section{Conclusions}
In this paper, we have studied the Gauss-Bonnet brane world
cosmology where the matter is confined to the brane through the
action of a confining potential, rendering the use of any junction
condition redundant. The modified Friedmann equations on the brane
were obtained and shown to be modified by various terms. As a
notable consequence, it was shown that dark energy can emerge as a
result of the extrinsic curvature in the Gauss-Bonnet brane world.
Therefore, this scenario features the possibility of an
accelerated expansion at the late stage of the cosmic evolution
without the need to invoke either a cosmological constant or a
quintessence component. It turns out that if the Gauss-Bonnet
coupling is positive, the universe undergoes an accelerated
expansion.  The addition of the Gauss-Bonnet term has resulted in
an estimate for the age of the universe that is compatible with
modern observational data, improving on our estimates in a
previous work where the same mechanism was used to localized the
matter on the brane in the absence of the Gauss-Bonnet term.


\begin{thebibliography}{99}
\bibitem{1} L. Randall and R. Sundrum,   {\it Phys. Rev. Lett.}
{\bf 83} (1999) 3370 [hep-ph/9905221].

\bibitem{2} L. Randall and R. Sundrum,   {\it Phys. Rev. Lett.}
{\bf 83} (1999) 4690 [hep-th/9906064].

\bibitem{3}  P. Brax and C. van de Bruck,
{\it Class. Quant. Grav.} {\bf 20} (2003) R201 [hep-th/0303095] ,\\
R. Maartens, Reference Frames and Gravitomagnetism, ed. J
Pascual-Sanchez et. al. (World Scientific,2001), p.93-119 [gr-qc/0101059],\\
D. Langlois,   {\it Prog. Theor. Phys. Suppl.} {\bf 148} (2003)
181 [hep-th/0209261].

\bibitem{4} B. Zwiebach, {\it Phys. Lett. } B {\bf 156} (1985) 315,\\ D. G. Boulware
and S. Deser, {\it Phys. Rev. Lett. } {\bf 55} (1985) 2656 .

\bibitem{5} A. H. Chamseddine, {\it Phys. Lett. } B {\bf 233} (1989) 291,\\ F. Muller-Hoissen,
{\it Nucl. Phys. } B {\bf 346} (1990) 235.

\bibitem{6} N. D. Birrell and P. C. W. Davies, Quantum Fields in Curved
Space (Cambridge University Press, Cambridge, England, 1982).

\bibitem{7} D. Lovelock,  {\it J. Math. Phys. } {\bf 12} (1971)
498,\\M. Farhoudi, {\it Gen. Rel. Grav.} {\bf 38} (2006) 1261
[physics/0509210].

\bibitem{8} J. E. Kim and H. M. Lee, {\it Nucl. Phys.} B {\bf 602}
(2001) 346 , {\it ibid} B {\bf 619} (2001) 763 [hep-th/0010093].

\bibitem{9} K. A. Meissner and M. Olechowski, {\it Phys. Rev. Lett.} {\bf 86}
(2001) 3708 [hep-th/0009122], {\it ibid} {\it Phys. Rev.} D {\bf
65} (2002) 064017 [hep-th/0106203].

\bibitem{10} Y. M. Cho and I. P. Neupane, {\it Int. J. Mod. Phys.} A {\bf 18}
(2003) 2703 [hep-th/0112227]
, \\
Y. M. Cho, I. P. Neupane and P. S. Wesson, {\it Nucl. Phys. } B
{\bf621} (2002) 388 [hep-th/0104227], \\
N. E. Mavromatos and J. Rizos, {\it Phys. Rev.} D {\bf 62} (2000)
124004 [hep-th/0008074], \\ I. P. Neupane, {\it JHEP} {\bf 0009}
(2001) 040 [hep-th/0008190], {\it ibid} {\it Phys. Lett.} B {\bf
512} (2001) 137  [hep-th/0104226].

\bibitem{11} N. Deruelle and M. Sasaki, {\it Prog. Theor. Phys.} {\bf 110} (2003) 441 [gr-qc/0306032].

\bibitem{12} B. Abdesselam and N. Mohammedi, {\it Phys. Rev. } D {\bf
65} (2002) 084018 [hep-th/0110143], \\ N. Deruelle and T. Dolezel,
{\it Phys. Rev. } D {\bf 62}
(2000) 103502 [gr-qc/0004021], \\
S. Nojiri and S. D. Odintsov, {\it JHEP } {\bf 0007}
(2000) 049 [hep-th/0006232],\\
J. E. Lidsey, S. Nojiri and S. D. Odintsov, {\it JHEP } {\bf 0206}
(2002) 026 [hep-th/0202198], \\
C. Charmousis and J. Dufaux, {\it Class. Quant. Grav. } {\bf 19}
(2002) 4671 [hep-th/0202107] ,\\
C. Germani and C. Sopuerta, {\it Phys. Rev. Lett. } {\bf 88}
(2002) 231101 [hep-th/0202060], \\ G. Kofinas, R. Maartens, and E.
Papantonopoulos, {\it JHEP } {\bf 0310} (2003) 066
[hep-th/0307138], \\ D. Iakubovskyi and Y. Shtanov, {\it Class.
Quant. Grav.} {\bf 22} (2005) 2415 [gr-qc/0408093].

\bibitem{13} W. Israel, {\it Nuovo Cim.} B {\bf 44} (1966) 1,
\\S. C. Davis, {\it Phys. Rev. } D {\bf 67} (2003) 024030 [hep-th/0208205],
\\E. Gravanis and S. Willison, {\it Phys. Lett. } B {\bf 562} (2003)
118 [hep-th/0209076].

\bibitem{14} R. A. Battye and B. Carter,  {\it Phys. Lett.} B {\bf 509}
(2001) 331  [hep-th/0101061].

\bibitem{15} V. A. Rubakov and M. E. Shaposhnikov, {\it Phys. Lett.} B
{\bf 125}, (1983) 136.

\bibitem{16} S. Jalazadeh and H. R. Sepangi, {\it Class. Quant. Grav.}
{\bf 22} (2005) 2035 [gr-qc/0408004].

\bibitem{17} M. Heydari-fard, M. Shirazi, S. Jalazadeh and H. R.
Sepangi, {\it Phys. Lett.} B {\bf 640}, (2006) 1 [gr-qc/0607067].

\bibitem{18} T. Shiromizu, K. Maeda and M. Sasaki, {\it Phys.
Rev.} D {\bf 62} (2000) 024012 [gr-qc/9910076].

\bibitem{20} K. Maeda, T. Torii, {\it Phys. Rev. } D {\bf 69} (2004)
024002 [hep-th/0309152].

\bibitem{21} M. D. Maia, E. M. Monte, J. M. F. Maia and J. S. Alcaniz,
{\it Class. Quant. Grav.} {\bf 22} (2005) 1623 [astro-ph/0403072].

\bibitem{maia1} M. D. Maia and E. M. Monte, {\it Phys. Lett.} A {\bf 297} (2002)
9 [hep-th/0110088].

\bibitem{22} M. D. Maia, E. M. Monte, J. M. F. Maia, {\it Phys. Lett.} B
{\bf 585} (2004) 11 [astro-ph/0208223].

\bibitem{23} M. Turner and M. White, {\it Phys. Rev. } D {\bf 56}
(1997) 4439 [astro-ph/9701138],\\ T. Chiba, N. Sugiyama and T.
Nakamura, {\it Mon. Not. R. Astron. Soc.} {\bf 289} (1997) L5
[astro-ph/9704199].

\bibitem{24} L. P. Eisenhart  1966 {\it Riemannian
Geometry}, (Princeton University Press).

\bibitem{cosmological constant} P. J. E. Peebles and
B. Ratra, {\it Rev. Mod. Phys.} {\bf 75} (2003) 559
[astro-ph/0207347],\\T. Padmanabhan, {\it Phys. Rept.} {\bf 380}
 (2003) 235 [hep-th/0212290].

\bibitem{quintessence} R. R. Caldwell, R. Dave and P. J. Steinhardt,
{\it Phys. Rev. Lett.}  {\bf 80} (1998) 1582 [astro-ph/9708069],\\
A. R. Liddle and R. J. Scherrer, {\it Phys. Rev.} D {\bf 59}
(1998) 023509 [astro-ph/9809272].


\bibitem{XCDM} Z. H. Zhu, M. K. Fujimoto and D.
Tatsumi, {\it Astron. Atrophys.} {\bf{372}} (2001) 377
[astro-ph/0107234],\\ P. S. Corasaniti, M. Kunz, D. Parkinson, E.
J. Copeland and B. A. Bassett, {\it Phys. Rev.} D {\bf70} (2004)
083006 [astro-ph/0406608].

\bibitem{25} M. H. Dehghani, {\it Phys. Rev. } D
{\bf 70} (2004) 064009 [hep-th/0404118].

\bibitem{26} D. G. Boulware and S. Deser, {\it Phys. Rev.
Lett.} {\bf 55} (1985) 2656.

\bibitem{27} L. M. Krauss and B. Chaboyer, Science {\bf 299}
(2003) 65 [astro-ph/0111597].

\bibitem{28} W. L. Freedman and M.S. Turner, {\it Rev. Mod. Phys }
{\bf 75} (2003) 1433 [astro-ph/0308418].


\bibitem{29} L. Knox, N. Christensen and C. Skordis, {\it Astrophys. J. }
{\bf 563} (2001) L95 [astro-ph/0109232].

\bibitem{30} W. Godlowski and M. Szydlowski, {\it
Gen. Rel. Grav.} {\bf 36} (2004) 767 [astro-ph/0404299],\\M. P.
Dabrowski, W. Godlowski and M. Szydlowski, {\it Mod. Phys.} D {\bf
13} (2004) 1669 [astro-ph/0210156].

\bibitem{31} G. Dvali, G. Gabadadze and M. Porrati, {\it Phys. Lett. } B {\bf 485} (2000)
208 [hep-th/0005016],\\G. Dvali and G. Gabadadze, {\it Phys. Rev.
} D {\bf 63} (2001) 065007 [hep-th/0008054].

\bibitem{32} C. Deffayet, {\it Phys. Lett. } B {\bf 502} (2001)
199 [hep-th/0010186].

\bibitem{33} C. Deffayet, G. Dvali and G. Gabadadze {\it Phys. Rev. } D
{\bf 65} (2002) 044023 [astro-ph/0105068].

\bibitem{34} J. S. Alcaniz, D. Jain and A. Dev, {\it Phys. Rev. } D {\bf 66} (2002)
067301 [astro-ph/0206448],\\J. S. Alcaniz, {\it Phys. Rev. } D
{\bf 65} (2002) 123514 [astro-ph/0202492].




\end{thebibliography}
\end{document}